\documentclass[aps,prl,amssymb,reprint,twocolumn,superscriptaddress,showpacs]{revtex4-1}

\usepackage[ansinew]{inputenc}
\usepackage{color}
\usepackage{graphicx}
\usepackage{amsmath}
\usepackage{color}

\begin{document}


\title{Ionization avalanching in clusters ignited by extreme-ultraviolet driven seed electrons}

\author{Bernd Schütte}
\email{schuette@mbi-berlin.de}
\affiliation{Max-Born-Institut, Max-Born-Strasse 2A, 12489 Berlin, Germany}
\affiliation{Department of Physics, Imperial College London, South Kensington Campus, SW7 2AZ London, United Kingdom}
\author{Mathias Arbeiter}
\affiliation{Institute of Physics, University of Rostock, Albert-Einstein-Str. 23, 18059 Rostock, Germany}
\author{Alexandre Mermillod-Blondin}
\affiliation{Max-Born-Institut, Max-Born-Strasse 2A, 12489 Berlin, Germany}
\author{Marc J. J. Vrakking}
\affiliation{Max-Born-Institut, Max-Born-Strasse 2A, 12489 Berlin, Germany}
\author{Arnaud Rouz\'{e}e}
\affiliation{Max-Born-Institut, Max-Born-Strasse 2A, 12489 Berlin, Germany}
\author{Thomas Fennel}
\email{thomas.fennel@uni-rostock.de}
\affiliation{Institute of Physics, University of Rostock, Albert-Einstein-Str. 23, 18059 Rostock, Germany}
\date{\today}

\begin{abstract}
We study the ionization dynamics of Ar clusters exposed to ultrashort near-infrared (NIR) laser pulses for intensities well below the threshold at which tunnel ionization ignites nanoplasma formation. We find that the emission of highly charged ions up to Ar$^{8+}$ can be switched on with unit contrast by generating only a few seed electrons with an ultrashort extreme ultraviolet (XUV) pulse prior to the NIR field. Molecular dynamics simulations can explain the experimental observations and predict a generic scenario where efficient heating via inverse bremsstrahlung and NIR avalanching is followed by resonant collective nanoplasma heating. The temporally and spatially well-controlled injection of the XUV seed electrons opens new routes for controlling avalanching and heating phenomena in nanostructures and solids, with implications for both fundamental and applied laser-matter science.
\end{abstract}

\maketitle

The strong-field ionization dynamics of solid-density matter exposed to NIR laser fields with intensities near the ionization threshold is a fundamental and challenging problem in laser-matter interactions. A process that is generic for this scenario is the generation of nanoscale plasmas via ionization avalanching, resulting from the strongly non-linear intertwining of collisional absorption and ionization. The quest for a quantitative microscopic understanding of this process is driven by its significance for a wealth of applications. These include laser-based material processing~\cite{englert08}, high harmonic generation (HHG)~\cite{ghimire11,schubert14,hohenleutner15,luu15,vampa15}, the realization of ultrafast lightwave electronics in dielectrics~\cite{schultze13} and the generation of shock waves~\cite{pezeril11,ecault13}.

In the last two decades, intense laser-cluster interactions have become an important platform for studying non-linear, collective, and correlated many-body processes in nanoplasmas for a wide spectral range from the NIR to the x-ray domain~\cite{saalmann06,fennel10}. A result that received particular attention was the unexpected highly charged ion emission following ionization of clusters by either an NIR~\cite{snyder96} or a vacuum-ultraviolet (VUV) laser pulse~\cite{wabnitz02}. By now it is well known that at long wavelengths, avalanching and transient resonant plasmon excitations are crucial to explain the observed highly efficient light absorption~\cite{ditmire96,zweiback99}, the emission of highly charged ions~\cite{snyder96,koeller99,zamith04,doeppner05} and the generation of fast electrons~\cite{shao96,springate03,saalmann08,fennel07b}. In clusters, the expansion of an initially overdense nanoplasma enables particularly strong resonant laser energy absorption, once the frequency of the collective electronic dipole mode $\omega_{\rm mie}~\propto\sqrt{\rho_{\rm ion}}$, the so-called Mie plasmon~\cite{mie08}, equals the laser frequency. Here $\rho_{\rm ion}$ is the ion charge density. While collision-mediated nanoplasma heating via inverse bremsstrahlung (IBS) was predicted to be important for excitation and ionization in the NIR~\cite{ditmire96} and VUV ranges~\cite{wabnitz02,santra03,siedschlag04}, vertical single-photon excitation of bound electrons becomes the key absorption mechanism in the XUV and in the x-ray regime~\cite{bostedt08,arbeiter10}.                                         
																						
An important concept for strong-field ionization dynamics of rare-gas clusters is the so-called ‘‘ionization ignition’’, proposed by Rose-Petruck \textit{et al.}~\cite{rose-petruck97}. Therein, the intensity threshold for efficient nanoplasma generation and heating was connected with the atomic ionization threshold, as the latter determines the seed electron generation required for avalanching. Direct evidence for this picture was found in a $z$-scan experiment on Xe clusters in He nanodroplets~\cite{doppner10}, where the sudden appearance of highly charged Xe ions was observed near the intensity threshold for tunnel ionization (TI) of atomic Xe. Evidence for the hypothesis that few seed electrons are sufficient for ignition was provided in an NIR few-cycle experiment on He nanodroplets, where weak doping with less than 10 Xe atoms was found to saturate the He$^{2+}$ emission at intensities well below the TI threshold of He~\cite{mikaberidze09, krishnan11}. In all these experiments, however, strong-field ionization by the NIR field itself was used to generated seed electrons.

In this Letter, we demonstrate an alternative concept that allows one to completely decouple the seed electron generation from the NIR driven ionization dynamics, both spatially and temporally. The main idea is to inject seed electrons via photoionization with a moderately intense ultrashort XUV pulse ($2\times 10^{10}$~W/cm$^2$) produced by HHG. The low intensity of the XUV pulse distinguishes our scenario from that of the theoretical study in~\cite{siedschlag05}, where strong cluster ionization was considered by a VUV pump pulse. We show that IBS heating, subsequent efficient NIR avalanching, and resonant excitation remain possible well below the TI threshold and can be triggered by just a few seed electrons. This fact is evidenced by the emission of highly charged ions up to Ar$^{8+}$ under conditions where no ion emission is observed without seeding. For NIR intensities as low as $3\times 10^{12}$~W/cm$^2$ we find ion emission up to Ar$^{4+}$, though the ponderomotive energy of $U_p=170$~meV is two orders of magnitude below the ionization potential of atomic Ar. The observations are well reproduced by molecular dynamics simulations. Our findings enable the study of low-intensity IBS during the early phases of avalanching and open a route to steer the spatial and temporal plasma formation in solids, with implications for laser material processing.


For the experiments, we use a Ti:Sapphire laser amplifier delivering pulses at 790~nm and operating at 50~Hz. The maximum energy achievable is 35~mJ for a pulse duration of 32~fs~\cite{gademann11}. The stretched laser beam is split by a beamsplitter into 2 parts that are individually compressed by two separate grating compressors. Up to 32~mJ of the pulse energy are used for HHG by focusing one beam with a 5~m long focal length spherical mirror into a 15~cm long gas cell statically filled with Kr. The NIR light used in the HHG process is blocked by a 100~nm thick Al filter after a propagation distance of 5~m. The second NIR laser pulse is collinearly overlapped with the XUV beam using a mirror with a 6~mm central hole. The maximum NIR pulse energy of 3~mJ can be reduced by the combination of a $\lambda$/2 waveplate and a polarizer. Both the XUV and NIR laser pulses are focused onto a cluster beam at the center of a velocity map imaging spectrometer~\cite{eppink97} by using a spherical multilayer mirror with a focal length of 75~mm. The cluster beam is generated by a piezoelectric valve operating at 10~Hz with a 0.5~mm diameter nozzle, and is placed 7~cm away from a 0.2~mm diameter skimmer. We estimate the average cluster sizes from the Hagena scaling law~\cite{hagena72}. Using the velocity map imaging spectrometer, we record the 2D projections of the electron and ion momentum distributions resulting from ionization by the two-color fields, from which we obtain kinetic energy spectra by using a standard Abel inversion procedure~\cite{vrakking01}.


\begin{figure}[tb]
 \centering
  \includegraphics[width=7cm]{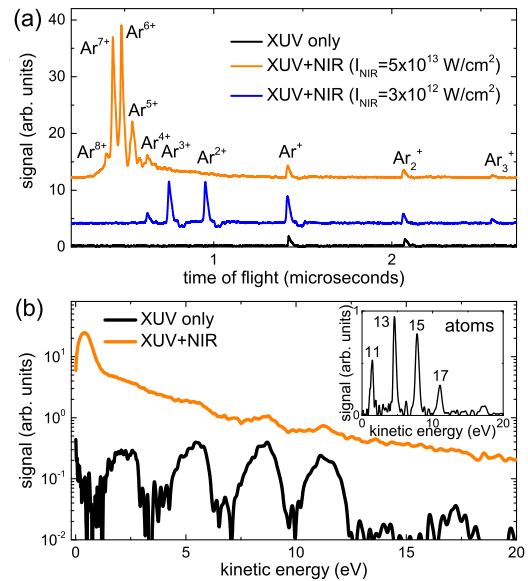}
 \caption{\label{figure1} (a) Ion TOF spectra from Ar clusters with an average size of $\langle N \rangle =3500$ atoms ionized by an XUV pulse only ($I_{XUV}=2\times 10^{10}$~W/cm$^2$) and with an additional NIR pulse at a time delay of 5~ps ($I_{NIR}=5\times 10^{13}$~W/cm$^2$ or $I_{NIR}=3\times 10^{12}$~W/cm$^2$). The different TOF spectra are plotted with a vertical offset. (b) The black curve shows the electron spectrum from clusters obtained by the XUV pulse only. An additional NIR pulse at a time delay of 600~fs and at a peak intensity of $5\times 10^{13}$~W/cm$^2$ (orange curve) strongly enhances the electron signal. The vertical axis has a logarithmic scale. The inset displays a photoelectron spectrum for the ionization of atomic Ar, with the main contributions coming from the 11th, 13th, 15th, 17th harmonics.}
\end{figure}

Ion time-of-flight (TOF) spectra resulting from the ionization of Ar$_N$ ($\langle N \rangle =3500$) by an XUV pulse and by the XUV-NIR pulse sequences are shown in Fig.~\ref{figure1}(a). The XUV intensity of $I=2\times 10^{10}$~W/cm$^2$ is two orders of magnitude lower than in our previous studies~\cite{schutte14a, schutte14b, schutte15a}. Ionization from the XUV pulse only (black curve) results in the observation of Ar$^+$ ions and larger ionic fragments such as dimers and trimers. When adding an NIR pulse with a peak intensity of $5\times 10^{13}$~W/cm$^2$ ($U_p=2.9~$eV) that is delayed by 5~ps (orange curve), ions with charge states up to Ar$^{8+}$ are generated. Remarkably, ion charges up to Ar$^{4+}$ are still observed when the NIR peak intensity is reduced by more than one order of magnitude to $3\times 10^{12}$~W/cm$^2$ (blue curve), corresponding to a ponderomotive potential of only 170~meV. Most importantly, no ion or electron signal from clusters was observed with the NIR pulse only in both cases, underlining the high contrast between seeded and unseeded excitation.

\begin{figure}
 \centering
  \includegraphics[width=8cm]{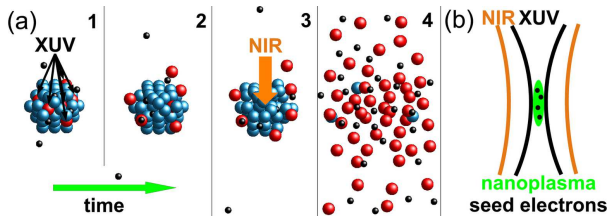}
 \caption{\label{figure2} (a) Scheme of the two-color cluster ionization processes, in which the cluster is ignited by a moderately intense XUV pulse in step (1), leading to the formation of a nanoplasma (step (2)). Neutral atoms are shown in blue, ions in red and electrons in black. In step (3), a time-delayed NIR pulse initially interacts with quasifree electrons and electrons that are weakly bound by atoms. Due to IBS heating, avalanching and resonance effects, the cluster is strongly ionized (step (4)). (b) Schematic of the NIR and XUV spatial profiles at the focus for Gaussian pulses. Since the XUV pulse (black) has a much smaller focus diameter than the NIR pulse (orange), focal volume averaging over different NIR intensities is avoided in the experiment. It is furthermore possible to restrict the ionization with NIR pulses to clusters where a nanoplasma is formed (green area), i.e. to the region of the highest XUV intensities.}
\end{figure}

Our observations can be explained by a modified ignition model (see Fig.~\ref{figure2}(a)). As shown in the photoelectron spectrum of atomic Ar in the inset of Fig.~\ref{figure1}(b), the XUV spectrum contains contributions from the 11th ($\hbar\omega=17.3$~eV), 13th (20.4~eV), 15th (23.6~eV) and 17th (26.7~eV) harmonics. Using atomic ionization cross sections for Ar~\cite{chan92}, we can estimate that less than 1~$\%$ of the atoms in the cluster are ionized at the applied XUV intensity of $I=2\times 10^{10}$~W/cm$^2$, see step (1) in Fig.~\ref{figure2}(a). In agreement with our earlier work~\cite{schutte14a} we find indications for frustrated recombination~\cite{fennel07} by observing very slow meV electrons in Fig.~\ref{figure1}(b). This suggests the onset of nanoplasma formation even at the moderate XUV intensity applied, cf. step (2) in Fig.~\ref{figure2}(a). Following ignition of the cluster ionization with the XUV pulse, the NIR laser pulse can interact with quasifree nanoplasma electrons and with electrons that are weakly bound by atoms, see step (3) in Fig.~\ref{figure2}(a) and Ref.~\cite{schutte15c}. For the chosen NIR pulse duration of 1~ps, electrons trapped within the clusters can be efficiently heated, resulting in extensive avalanching and strong ionization, which triggers cluster expansion (step 4 in Fig.~\ref{figure2}(a)). Strong NIR driven ionization is also supported by the fact that the electron signal from Ar clusters (Fig.~\ref{figure1}(b)) is increased by up to two orders of magnitude when adding the NIR pulse ($I=5\times 10^{13}$~W/cm$^2$).

In recent x-ray laser-cluster experiments it was demonstrated that the effects of focal volume averaging and the cluster size distribution crucially affect the measured ion spectra and need to be eliminated to measure meaningful ion charge state distributions at high x-ray intensity~\cite{gorkhover12}. A major advantage of our scheme is that the volume averaging over the NIR laser focus can be avoided via the spatially selective seed electron generation. As the NIR beam has a significantly larger focal spot size ($\approx 12~\mu$m) than the XUV beam ($\approx 3~\mu$m), seeded avalanching and nanoplasma formation is restricted to the center of the NIR beam where the NIR intensities are close to the peak value (see Fig.~\ref{figure2}(b)). Furthermore, NIR-induced electron and ion emission is restricted to larger clusters that experience the XUV peak intensity as the number of seed electrons per cluster scales with the number of cluster atoms $N$ and will drop below unity for small clusters / lower XUV intensity. The resulting selectivity makes the presented scheme suitable for a detailed comparison of experiment and theory.
																																															
\begin{figure}[tb]
 \centering
  \includegraphics[width=7cm]{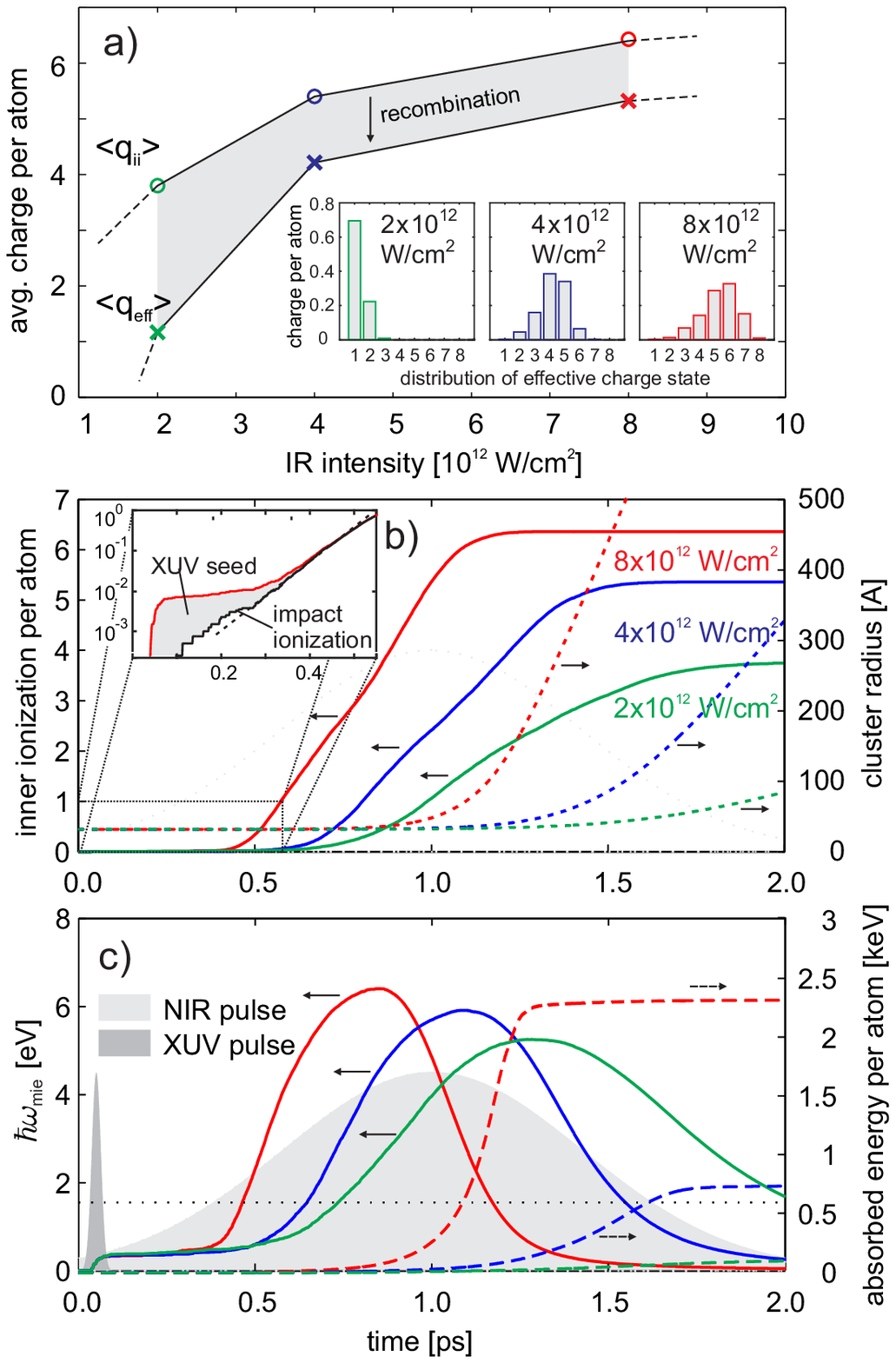}
 \caption{\label{figure_sim} Molecular dynamics simulations on Ar$_{3871}$. A 30~fs XUV seed pulse at $\hbar \omega=20~{\rm eV}$ ($I_{\rm XUV}=2.5\times 10^{10}~{\rm W/cm^2}$) is followed by a 1~ps NIR pulse (delayed by $\approx 1$~ps) at 800~nm (intensities as indicated). (a) Average inner charge state, $\langle q_{\rm ii}\rangle$, and predicted final effective charge state $\langle q_{\rm eff}\rangle$ by taking into account electron-ion recombination (gray area). Insets show corresponding simulated final effective ion charge spectra. (b) Temporal evolution of inner ionization (solid curves) and root-mean-square cluster radius (dashed curves); the inset shows the evolution of inner ionization during the seeding step and the subsequent avalanching process for the highest intensity on a logarithmic scale. The black dashed curve corresponds to an exponential growth. (c) Dynamics of the predicted Mie frequency $\hbar\omega_{\rm mie}$ of the nanoplasma (solid curves) and evolution of the total energy absorption (dashed curves). The gray areas indicate the intensity envelopes of the XUV and NIR fields. The dashed-dotted horizontal line corresponds to the NIR photon energy.}
\end{figure}

In order to analyze the evolution of the ionization and avalanching processes, we have performed semiclassical molecular dynamics simulations~\cite{arbeiter14} for parameters similar to the measurements, see Fig.~\ref{figure_sim}. In agreement with our experimental results, the predicted final ion charge spectra show high ionization stages up to Ar$^{8+}$ (Fig.~\ref{figure_sim}(a)) for the seeded case. The average final ion charge states increase rapidly up to NIR intensities of about $4\times10^{12}$~W/cm$^2$ before they begin to level out. The stages of the seeded avalanching process and the highly charged ion generation can be understood from the time-resolved analysis of the simulations in Figs.~\ref{figure_sim}(b)-(c). After the generation of only a few seed electrons, an exponential increase of the inner ionization (electrons removed from host atoms), $\langle q_{ii}\rangle$, in the leading edge of the NIR pulse documents the self-amplified character of the avalanche, resulting in the generation of (quasi)free electrons in each optical half cycle that contribute to the generation of even more electrons in the next cycles via impact ionization (see inset of Fig.~\ref{figure_sim}(b)). Note that values of $\langle q_{ii}\rangle>3$ are achieved for all investigated cases and that cluster expansion begins near the NIR peak intensity (Fig.~\ref{figure_sim}(b)). It has to be emphasized that non-resonant IBS heating is responsible for energy absorption in the early phase, while strong resonance heating via collective electron excitations becomes increasingly important in later stages. The latter process is linked to the Mie plasmon, which emerges for sub-wavelength particles and describes the collective oscillation of electrons with respect to the ions. The increase of the charge density via inner ionization and its decrease through cluster expansion lead to a time-dependent energy of the Mie plasmon (Fig.~\ref{figure_sim}(c)), where the resonance condition $\omega_{\rm NIR}=\omega_{\rm mie}$ is met twice. While the ionization driven first resonance is unimportant for the total energy absorption, the expansion induced resonance largely governs the total energy absorption~\cite{fennel10}. The fact that the expansion driven resonance occurs late in the trailing edge of the NIR pulse for the lowest intensity explains the stronger effect of recombination in this case (cf. Fig.~\ref{figure_sim}(a)), since recombination is more efficient for lower electron temperature~\cite{hahn97}.

\begin{figure}
 \centering
  \includegraphics[width=7cm]{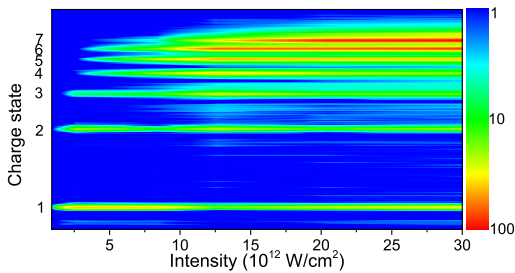}
 \caption{\label{figure4} Ion charge-state distributions from Ar clusters with $\langle N\rangle=18000$ atoms at different NIR intensities. An XUV pulse ($I=2\times 10^{10}$~W/cm$^2$) preceeds the NIR pulse by 5~ps. The average charge state increases as a function of the NIR intensity and is peaked at Ar$^{7+}$ for an intensity of $3\times 10^{13}$~W/cm$^2$. The ion signal is shown in a logarithmic scale.}
\end{figure}

The simulations show a fast increase of the ionization states as function of the NIR laser intensity and indicate the onset of saturation for intensities well below the TI threshold for Ar. This behavior is also observed in the experimental ion charge spectra shown for different intensities in Fig.~\ref{figure4}. We can furthermore conclude from the simulations that, under the given experimental conditions, single-pulse excitation, both NIR-only as well as XUV-only, does not significantly ionize the system ($q^{\rm XUV}_{\rm avg}<0.01$ and $q^{\rm NIR}_{\rm avg}=0$), highlighting the synergetic action of XUV induced seeding and long-wavelength driven avalanching and nanoplasma heating.

As multiphoton ionization is avoided in our experiment, we can conclude that the efficient absorption of laser energy in early stages, i.e. far away from the plasmon resonance condition, can only be explained by inverse bremsstrahlung (IBS), in agreement with our simulations. Therefore, our experimental results confirm that IBS is efficient in rare-gas clusters at ponderomotive potentials on the order of 100~meV. This shows that IBS may indeed play an important role in the ionization of clusters with intense VUV pulses, where similar ponderomotive potentials were applied~\cite{wabnitz02,santra03,siedschlag04}. In the future, our ignition method is expected to enable the quantitative study of IBS at low ponderomotive potentials, at low plasma densities, and as function of wavelength. It should be applicable even in the VUV range, as long as VUV single-photon ionization can be excluded by choosing sufficiently small VUV photon energies.

Our concept of cluster ignition using an ultrashort XUV pulse is very versatile and is expected to be applicable in a wide range of laser parameters to study and control fundamental processes of light-matter interactions. Using attosecond XUV pulses would allow the investigation of the plasma dynamics on a sub-NIR-cycle timescale. In the context of HHG in solids~\cite{ghimire11,schubert14,hohenleutner15,luu15,vampa15}, our approach could be used to temporally control the HHG process in order to better understand and possibly avoid the undesired damaging effects of plasma formation. Applied to the surface ablation of semiconductors and dielectrics, an irradiation sequence composed of an (X)UV laser pulse followed by an NIR laser pulse offers a higher micromachining precision and a higher efficiency than conventional irradiation schemes involving a single color or existing dual beam methods~\cite{theberge05,zoppel07,lin10,grojo10}. First, the interaction footprint is determined by the size of the (X)UV spot only, leading to a dramatic improvement of the spatial precision. Furthermore, the overall efficiency of the process (i.e. the ratio of the energy required to evaporate the volume removed to the energy contained in the laser electromagnetic field) may significantly increase because the use of a low-intensity (X)UV pulse allows one to apply an NIR pulse with a much lower intensity than in conventional schemes. 


In summary, we have reported on a novel concept to ignite NIR-driven ionization of solid-density matter by generating seed electrons using an ultrashort XUV pulse. High charges up to Ar$^{4+}$ were generated from clusters with an NIR pulse at an intensity of only $3\times 10^{12}$~W/cm$^2$, i.e. far below the tunnel ionization threshold. From a combination of experimental and theoretical studies, we could conclude that IBS and avalanching play a key role for the initial charging of the clusters, which is strongly enhanced at the plasmon resonance, leading to the observed high charge states. Our results demonstrate for the first time that inverse bremsstrahlung plays an important role in rare-gas clusters interacting with intense laser pulses, which have a ponderomotive potential as low as 100~meV.

\begin{acknowledgments}
We would like to thank M. Yu. Ivanov for fruitful discussions. B. S. is grateful for funding from the DFG via a research fellowship. M.A. and T.F. gratefully acknowledge financial support from the DFG within SFB652 and computer time provided by the North-German Supercomputing Alliance (HLRN, project mvp00010).
\end{acknowledgments}


%

\end{document}